\documentclass[%
 reprint,
 amsmath,amssymb,
 aps,
 prk,
floatfix,
]{revtex4-2}

\usepackage{graphicx}
\usepackage{dcolumn}
\usepackage{bm}
\usepackage{hyperref}

\usepackage[ruled,vlined]{algorithm2e}

\makeatletter
\newcommand*{\balancecolsandclearpage}{%
  \close@column@grid
  \clearpage
  \twocolumngrid
}
\makeatother

\begin{document}

\preprint{APS/123-QED}

\title{Historical Contingencies Steer the Topology of Randomly Assembled Graphs} 



\author{Cole Mathis\textsuperscript{\textdagger}}
 \email{cole.mathis@asu.edu}
\affiliation{
 Biodesign Institute, Arizona State University, Tempe, AZ, USA
 }
\affiliation{
 School of Complex Adaptive Systems, Arizona State University, Tempe, AZ, USA 
}


\author{Harrison B. Smith\textsuperscript{\textdagger}}
\affiliation{
 Earth-Life Science Institute, Institute of Science Tokyo, Meguro-ku, Tokyo, Japan
}%
\affiliation{
 Blue Marble Space Institute of Science, Seattle, WA, USA
}%
\thanks{These authors contributed equally}

\date{\today}

\begin{abstract}
Graphs are used to represent and analyze data in physics, biology, chemistry, planetary science, and the social sciences. Across domains, random graph models relate generative processes to expected graph properties, and allow for sampling from distinct ensembles. Here we introduce a new random graph model, inspired by assembly theory, and characterize the graphs it produces. We show that graphs generated using our method represent a diverse ensemble, characterized by a broad range of summary statistics, unexpected even in graphs with identical degree sequences. Finally we demonstrate that the distinct properties of these graphs are enabled by historical contingencies during the generative process. These results lay the foundation for further development of novel sampling methods based on assembly theory with applications to drug discovery and materials science.
\end{abstract}

\maketitle

\textit{\label{sec:intro}Introduction---}Many physical systems are combinatorial in character, particularly in material science and chemistry~\cite{polya2012combinatorial}. In many situations the possible configurations of these systems are vast, and it is impractical, or impossible, to exhaustively enumerate them~\cite{polishchuk2013estimation, choubisa2023accelerated, restrepo2022chemical, coley2021defining}. For example, drug candidates cannot be found directly in small molecule space: enumerating all possible small molecules with the appropriate properties is typically intractable~\cite{polishchuk2013estimation}. Often it is sufficient to generate statistical samples from these systems~\cite{choubisa2023accelerated, coley2021defining}. However, in even the simplest cases it is often impossible to define procedures that sample uniformly from the space of all possible combinations. A famous example is provided by Bertrand's Paradox, which illustrates three different procedures to sample chords from a circle, each yielding different distributions despite all being apparently equivalently random ~\cite{shackel2007bertrand}. Thus, sampling procedures over combinatorial systems define a characteristic ensemble which describes the entities they are likely to produce~\cite{erdos1960evolution, albert2002statistical}. As such, they must be designed strategically to effectively explore combinatorial spaces~\cite{gomes2006near, aerts2014solving}. 

Graphs (or networks) are inherently combinatorial structures~\cite{erdos1960evolution, albert2002statistical}. They are a simple representation of relationships between components, allowing them to represent diverse physical structures such as molecules and materials, as well as biological and social systems~\cite{polya2012combinatorial, newman2018networks}. The most common sampling procedure for graphs is the Erdős-Renyi (ER) random graph model, in which edges are placed independently between N nodes with probability p, or alternatively M edges are assigned between N nodes uniformly and at random~\cite{erdos1960evolution}. This model generates networks with a Poisson degree distribution and uncorrelated edges~\cite{newman2018networks}. Real-world networks often exhibit features that ER graphs do not, such as heavy-tailed degree distributions, high clustering coefficients, and small-world properties~\cite{albert2002statistical}. The Watts-Strogatz model was introduced to address the latter two features by rewiring a regular lattice, preserving clustering while reducing path lengths~\cite{newman2002random}. The Barabási-Albert (BA) model, based on growth and preferential attachment, generates power-law degree distributions, capturing the ``rich-get-richer" mechanism seen in many empirical networks~\cite{albert2002statistical}. The Kronecker Graph model uses a recursive approach to generate networks, by iteratively applying the Kronecker product to an initial adjacency matrix~\cite{leskovec2010kronecker}. A stochastic version of this can be used to sample random graphs from the deterministic recursive process, and the parameters of this model can be fit to large real world networks~\cite{leskovec2010kronecker}. Other random graph models have used a recursive approach to generate scale-free or fractal-like networks (e.g.~\cite{ravasz2003hierarchical}).

\begin{figure*}[!ht]
\centering
\includegraphics[width=\textwidth]{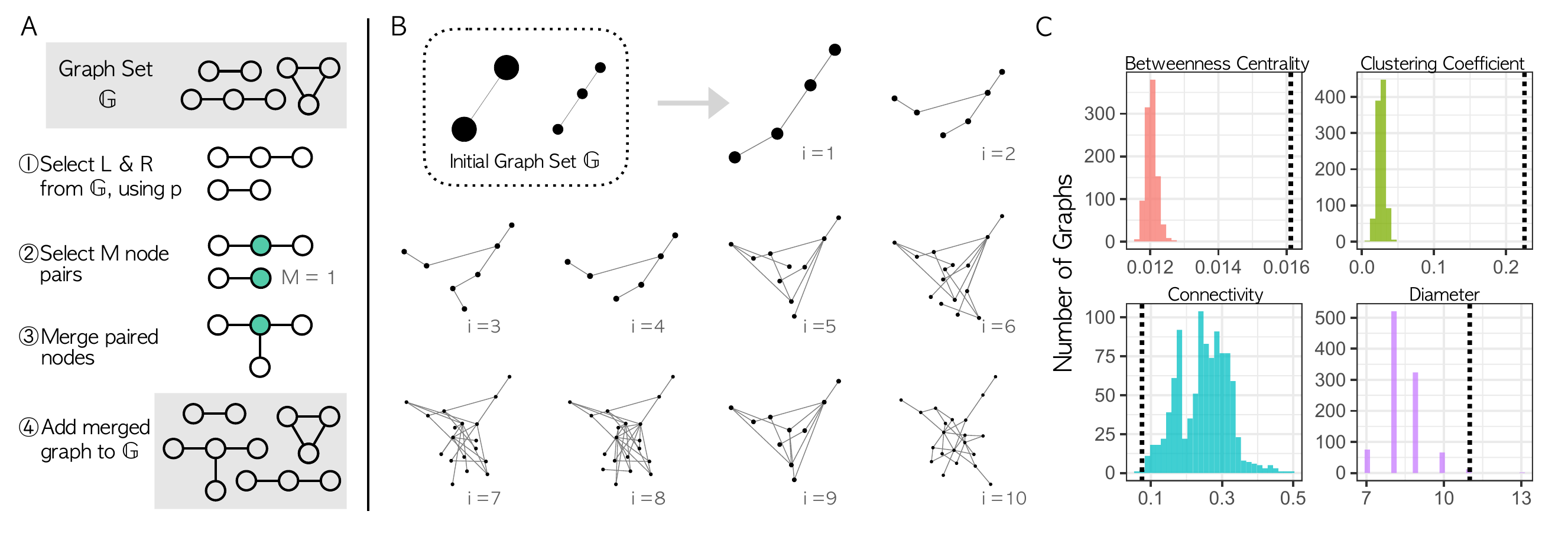}
\caption{\label{fig:concept}\textbf{Generation Algorithm Figure and Degree Trajectories}. \textbf{(a)} The algorithm starts with a multiset of graphs. Two graphs are randomly chosen to be merged, and are then merged on one or more pairs of nodes (one node from each graph). Effectively, these nodes become the same node. The resulting graph is added back into the multiset of graphs, and the process is repeated. \textbf{(b)} A specific trajectory of the algorithm for 10 iterations, showing the initial graphs and the first 10 generated graphs. \textbf{(c)} A randomly assembled graph was constructed using 25 iterations, the betweenness centrality, clustering coefficient, connectivity, and diameter were calculated (black dashed lines), 1000 random graphs with the same degree sequence as the assembled graph were generated and the histograms show their corresponding statistics.}
\end{figure*}

Here we introduce a new generative procedure for sampling graphs, based on assembly theory, and characterize the ensemble it produces. Assembly theory (AT) is a new theoretical framework for characterizing selection across diverse objects, most importantly in molecules~\cite{sharma2023assembly}. In AT, objects are defined as finite in extent, persistent in time, distinguishable, and decomposable into basic building blocks. Graphs satisfy all of these conditions (with the possible exception of persistence in time, as the ontological status of mathematical entities is debated). A central quantity in AT is the assembly index of objects, which is the minimum number of joining operations required to construct the object from basic components, in which recursively generated objects can be reused as a single joining operation~\cite{sharma2023assembly, marshall2022formalising}. The assembly index can be measured empirically for molecules, enabling novel approaches to quantify life detection, evolutionary relationships, and material characteristics~\cite{marshall2021identifying, jirasek2024investigating, kahana2024constructing, patarroyo2025quantifying}. The definition of the assembly index implies a constructive procedure that when paired with empirical data can produce structures with desirable properties, such as similarity to target compounds or increased ``drug-likeness"~\cite{liu2021exploring, pagel2024mapping}. Here we generalize this approach to graphs by defining the constructive steps, and then characterize the ensemble of graphs induced by this procedure. This procedure is distinctive from other random graph models because it both recursively reuses substructures (without preserving the entire adjacency matrix) and is dynamic in nature~\cite{leskovec2010kronecker, newman2002random}. Here we describe the sampling procedure in detail, provide an algorithmic implementation~\cite{AssemblingGraphs2025}, and show that the graphs generated by this procedure are exceptional (compared to graphs with identical degree sequences) based on their global topological properties such as mean betweenness, clustering coefficient, and algebraic connectivity~\cite{fiedler1973algebraic, newman2018networks}. Finally we demonstrate how the contingency induced by this procedure enables the generation of diverse samples with these exceptional properties, and discuss the implications to exploration of chemical space.

\textit{\label{sec:model}Model Description---}The random graph assembly process begins with a multiset $\mathbb{G}$ containing simple graphs. At each iteration, two graphs are selected from $\mathbb{G}$, with replacement. The first graph, $L$, is chosen through a biased selection process: with probability $p$, $L$ is selected uniformly at random from $\mathbb{G}_{\max}$ (a subset of $\mathbb{G}$ containing only graphs with the most nodes); with probability $1-p$, $L$ is selected uniformly at random from $\mathbb{G}$. The second graph, $R$, is always selected uniformly at random from $\mathbb{G}$. It is possible for $L$ and $R$ to be identical graphs. The process then attempts to merge $M$ pairs of vertices between $L$ and $R$, with a vertex pair always containing one node in $L$ and one node in $R$. No vertex may be in more than one pair. When two vertices are merged, the resulting vertex inherits all edges from both original vertices, while maintaining the properties of a simple graph: parallel edges are combined, and self-loops are prohibited. The resulting graph is then added to $\mathbb{G}$. This entire process is repeated for $N$ iterations. At the end of the run $\mathbb{G}$ contains a variety of randomly generated graphs. We focus our analysis on the largest graph in the set at each iteration, though the properties of the entire set are an interesting topic for future study. \textbf{Fig. \ref{fig:concept}} illustrates this procedure conceptually (panel \textbf{a}), and gives an example trajectory of largest graphs assembled for 10 iterations of this process (panel \textbf{b}).

Specifying an instance of this algorithm requires specifying an initial multiset of graphs $\mathbb{G}$, the number of iterations $N$, the probability $p$ of selecting $L$ from $\mathbb{G}_{\max}$, and a procedure for selecting $M$. For the results presented here we always initialize $\mathbb{G}$ with a path graph of two nodes and a path graph of three nodes (see Appendix \ref{si:algorithm}). We chose $M$ at each iteration uniformly at random from the discrete range $[1,m]$. Higher values of $p$ yield larger graphs in fewer iterations but reduce the diversity of accessible graphs.

\textit{\label{sec:results}Results---}To characterize the graphs generated by this algorithm we generated trajectories for a variety of inputs, most importantly varying the range of $M$, from 1 to 6, and for varying values of $p$.  If $M$ is always 1, the resulting graphs are always tree-like graphs and converge to a mean degree of 2. If $M \in [1,2]$ the graphs can generate a diverse set of outcomes, and each trajectory can yield graphs with varying mean degrees depending on early fluctuations in the population $\mathbb{G}$. The same is true for $M \in [1,3]$ and for larger ranges. 

To further characterize the produced graphs, we computed several global properties for the largest graphs from the trajectories. Specifically we computed the global clustering coefficient, the mean betweenness centrality, diameter, and algebraic connectivity of the graphs. To evaluate these we compared these graph measures to the corresponding values for ER random graphs with the same number of nodes and edges, as well as random configuration model graphs with the same degree sequences. The ER random graphs provide a control for the expected values based purely on the node and edge counts, while the configuration model graphs yield insights into the properties of the graph that are exceptional even when controlling for the degree sequence. In the event these randomizations produced disconnected graphs, we compared the network measures to those of the largest connected component. The results of one such randomization with a single assembled graph, and 1000 random configuration graphs is shown in \textbf{Fig. \ref{fig:concept} C}. The randomly assembled graph in this case exhibits a high betweenness centrality, clustering coefficient and diameter, but a low algebraic connectivity compared to random graphs of the same degree sequence. The measures suggest this graph contained many triangles, and several central nodes that could be removed to easily fracture the network. 

\begin{figure}
\centering 
\includegraphics[width=75mm]{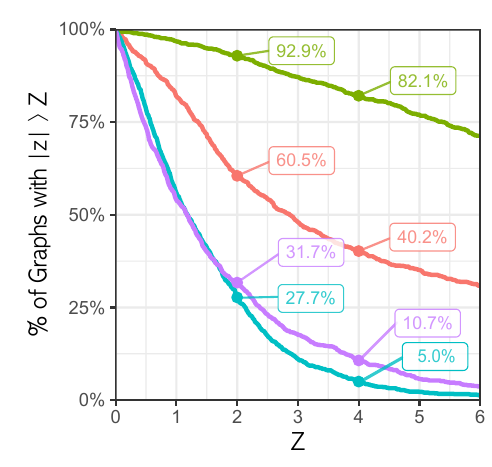}
\caption{\label{fig:mixing}\textbf{Randomly assembled graphs exhibit properties atypical of their degree sequence} For each of 1000 randomly assembled graphs generated by 25 iterations of the assembly algorithm, 1000 random configuration graphs were sampled using the original graph's degree sequence. The Z-score for each randomly assembled graph's topological properties was calculated relative to these sampled graphs. The plot shows the percentage of graphs exceeding given absolute Z-score thresholds ($|Z|$). Green represents clustering coefficient, red denotes mean betweenness, purple indicates diameter, and blue corresponds to algebraic connectivity (as in Fig. \ref{fig:concept}C). Callouts indicate the percentage of graphs exceeding thresholds $|Z| = 2$ and $|Z| = 4$. }
\vspace{-5pt}
\end{figure}

We repeated this analysis for 1000 randomly assembled graphs, calculating these parameters after 25 iterations, using 1000 controls for each graph. We found that the assembled graphs frequently exhibited values in the most extreme upper/lower tails of these parameters as compared to randomized configuration models with the same degree sequences. To quantify this we calculated the $Z$ score of the relevant statistic for randomly assembled graph compared in the ensemble of the random configuration graphs derived from it. The results are shown in Fig. \ref{fig:mixing} ($p=0.5$, and $M \in [1,3]$). These results demonstrate that our algorithm can sample graphs which include typical random graphs but often represent extremely atypical graphs, indicating the algorithm is consistently sampling graphs that represent extreme cases compared to other sampling procedures.  

\begin{figure}
\centering
\includegraphics[width=75mm]{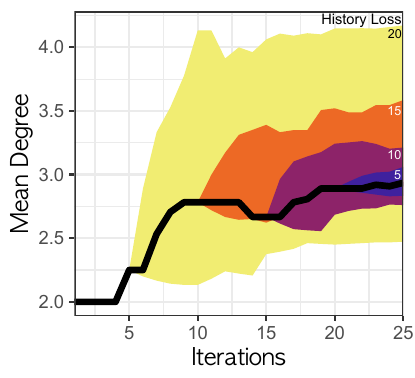} 
\caption{\label{fig:history}\textbf{Historical contingency allows randomly assembled graphs to sample diverse topologies.} The mean degree as a function of iteration, grouped by the step at which the alternative trajectories diverge (history loss). When alternative trajectories diverge from the reference trajectory, they initially produce graphs which have similar mean degrees to the reference trajectory, but as they continue on their own path, the similarity with the reference graph mean degree diverges. The cones of color indicate the maximum and minimum mean degree observed at each iteration, within that history loss category. This plot emphasizes that the model we use for random graph generation does not always produce similar graphs, even when using the same starting set of graphs and same set of parameters.
}
\vspace{-5pt}
\end{figure}

A key feature of this algorithm is the contingency that is induced within a single trajectory. To explore how this feature controls the properties of the produced graphs we performed simulations to resample from trajectories with some of this contingency removed in two different ways. First we explored the effect of resampling the trajectories to understand how the historical features controlled the entire ensemble of possible trajectories. Specifically we do the following: for a single trajectory of the algorithm (a reference trajectory), we generate a new input multiset, $\mathbb{R}'_c$, which is the final multiset of graphs, $\mathbb{R}$, accumulated in the reference trajectory with the graphs accumulated in the last $c$ steps removed. We then run the algorithm with $\mathbb{R}'_c$ for $c$ steps, yielding a new trajectory with an identical history for the first $N-c$ steps, but allowing for a distinct final $c$ steps. We refer to this trajectory as having a \textit{history loss} of $c$ steps. \textbf{Fig. \ref{fig:history}} shows the distribution of mean degrees for 1000 trajectories for different values of $c$ ($p$ = 0.5, $M \in [1:2]$). The intervals indicate the inter-quantile range between 1\% and 99\% for the distributions. Larger values of $c$ enable larger variations in the final properties of the graph, consistent with the idea that fluctuations in the trajectory enable sampling of distinct sections of the graph space later in the trajectory. 

\begin{figure}
\centering
\vspace{10pt}
\includegraphics[width=80mm]{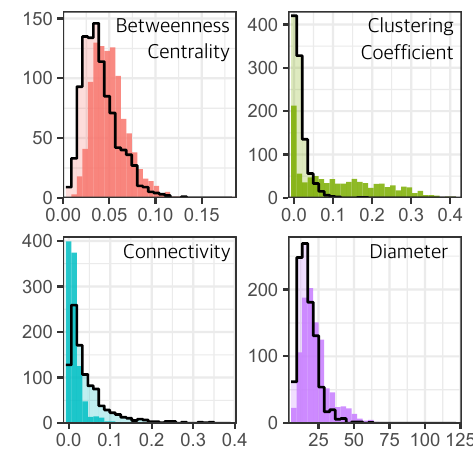}
\caption{\label{fig:history_control}\textbf{Topological properties of randomly assembled graphs depend on history and dynamics.} The distribution of graph topological properties is shown for each property for the randomly assembled graphs, and for the control that randomizes connectivity (preserving number of nodes and edges) at each step after selecting graphs to merge. This controls for the size of the joined graphs and the sequence of joining operations. The distributions for the original randomly assembled graphs (shown in opaque colors) and control graphs (outlined in black) are distinct, indicating that the historical dependence of graph reuse is partially responsible for the observed topological properties. 
}
\end{figure}

Next we explored the effect of removing the influence of previous graphs on the produced graph—effectively erasing history within a trajectory—while controlling for the sequential construction process (including the controlling for the mean degree). Specifically we generated a reference trajectory, tracking the size (number of nodes and edges) of each pair of graphs that was joined. Then we run a variation of the algorithm where after $L$ \& $R$ are chosen their connectivity is randomized, such that the number of nodes and edges are preserved and the graph is still connected, but otherwise random (see Appendix \ref{si:algorithm}). \textbf{Fig. \ref{fig:history_control}} shows the difference between our algorithm and this random control for the global measures we calculated with the original graphs in solid colors, and the randomized controls shown in black. The distributions of graph properties for the original and randomized methods are all statistically significant.

\textit{\label{sec:conclusion}Conclusion---}Graphs are widely used mathematical abstractions across various physical and biological sciences. However, their analysis has primarily emphasized large, complex graphs typical of social or biological systems, which are difficult to measure directly. Here, we introduced a novel algorithm based on assembly theory for sampling random networks. We demonstrated that this approach enables sampling networks with diverse properties, yielding structures with exceptional global characteristics atypical of their degree sequences. We further showed that the algorithm's historically contingent features drive these key properties.

Earlier work has applied related, more narrowly focused ideas from assembly theory to sample molecules with enhanced drug-like properties and to explore chemical spaces around known natural products~\cite{liu2021exploring,pagel2024mapping}. Our algorithm formalizes these earlier approaches, highlighting two critical yet underexplored parameters, $p$ and $M$, and extends their application from molecular fragments to graphs. We anticipate this method could be integrated with existing techniques, such as genetic algorithms~\cite{spiegel2020autogrow4, verhellen2020illuminating}, by generating distinctive molecules rare in other sampling schemes, suitable for subsequent optimization for drug discovery and materials science. Although we primarily addressed the algorithmic implementation, our empirical findings may inform further foundational work in assembly theory and the design of algorithms to compute assembly indices~\cite{marshall2022formalising, seet2024rapid, flamm2025assembly}.

\vspace{15pt}

\begin{acknowledgments}
We wish to acknowledge the support of the World Research Hub (WRH) program at Institute of Science Tokyo for graciously supporting the travel and lodging funding necessary to complete this project. CM acknowledges helpful conversations with Swanand Khanapurkar, Dániel Czégel, and Sara Imari Walker.

\end{acknowledgments}

\vspace{15pt} 
\noindent \textbf{Code Availability}. Code associated with this manuscript is available at \url{https://github.com/mathis-group/AssemblingGraphs.jl} 

\balancecolsandclearpage
\bibliography{main}

\balancecolsandclearpage
\appendix

\section{Graph Assembly Algorithm Details}
\label{si:algorithm}
\SetKwComment{Comment}{ \# }{}

\begin{algorithm}[h]
\caption{Graph Assembly Algorithm}
\KwData{$p \in [0,1]$ (prob. of forcing largest graph selection), $M \in \mathbb{N}$ (desired node merges), $N \in \mathbb{N}$ (iterations)}
\KwResult{Set of assembled graphs $\mathbb{G}$}
\textbf{Notation:}
\begin{itemize}
\item $V(G)$ -- vertex set of graph $G$
\item $H(v)$ -- neighborhood of vertex $v$
\item $U(S)$ -- uniform random selection from $S$
\item $P_n$ -- path graph with $n$ nodes
\end{itemize}
\textbf{Initialize:}\;
$\mathbb{G} \leftarrow \{\{P_2, P_3\}\}$

\For{$i = 1$ \KwTo $N$}{
   \textbf{Select graph $L$}\;
   \eIf{$U([0,1)]) < p$}{
       $L \leftarrow \arg\max\{|V(G)| : G \in \mathbb{G}\}$\;
   }{
       $L \leftarrow U(\mathbb{G})$
   }
   
   \textbf{Select graph $R$} $\leftarrow U(\mathbb{G})$\;
   $m \leftarrow \min(M, \min(|V(L)|, |V(R)|))$\;
   
   \textbf{Select $m$ vertex pairs} $\{(v_1^L, v_1^R), \ldots, (v_m^L, v_m^R)\}$ where\;
   \Indp $v_i^L \in V(L)$ \& $v_i^R \in V(R)$ for all $i \in \{1, \ldots, m\}$\;
   $v_i^L \neq v_j^L$ \& $v_i^R \neq v_j^R$ for all $i \neq j$\;
   \Indm
   
   \ForEach{pair $(v_i^L, v_i^R)$}{
       \textbf{Create merged vertex} $v_{merged}$:\;
       \Indp $H(v_{merged}) \leftarrow H(v_i^L) \cup H(v_i^R)$\;
       \textbf{Remove parallel edges and self-loops}\;
       \Indm
   }
   
   $G_{merged} \leftarrow$ resulting graph after all $m$ merges\;
   $\mathbb{G} \leftarrow \mathbb{G} \cup \{G_{merged}\}$\;
}
\end{algorithm}

\section{Number of nodes merged, $M$.} The actual number of merges performed is limited by the sizes of the selected graphs---specifically, the number of merges is at minimum the smaller of the two graphs' vertex counts, $m = \min(|V(L)|, |V(R)|)$. We chose $M$ at each iteration uniformly at random from the discrete range $[1,m]$.

\section{ Choosing the initial $\mathbb{G}$.} Our graph assembly requires combining graphs with at least 1 edge to allow the possibility of the resulting graph to be larger than the two merged graphs (e.g., consider the scenario with a singleton node graph $L_{\text{singleton}}$. Because merging two graphs requires $M \geq 1$, joining $L_{\text{singleton}}$ with any graph $R$ will always result in $R$). While the simplest such set of initial graphs would be two path graphs of two nodes each, this will always lead to a path graph of three nodes before complexifying further, hence we initialize our algorithm with two path graphs, of three nodes and two nodes respectively. We anticipate future work could consider optimization of this initial set to sample different ensembles of assembled graphs

\section{ Choosing the largest graphs.} When multiple graphs in $\mathbb{G}$ meet the condition for graph with the largest graph (by number of nodes), one is chosen uniformly at random.





\end{document}